\documentclass[11pt]{article}
\usepackage{epsfig,sint,cite,amsfonts}
\usepackage{amssymb}
\usepackage{amsmath}
\usepackage[english]{babel}
\usepackage{simplewick}
\usepackage{bm}
\usepackage{graphicx}

\renewcommand{\vec}[1]{\boldsymbol{#1}}
\newcommand{\be}{\begin{equation}}
\newcommand{\ee}{\end{equation}}
\newcommand{\ba}{\begin{eqnarray}}
\newcommand{\ea}{\end{eqnarray}}
\newcommand{\la}{\label}

\newcommand{\<}{\langle}
\renewcommand{\>}{\rangle}

\begin{document}

\begin{flushright}
MITP/18-054
\end{flushright}
$ $ \bigskip\bigskip\bigskip

\begin{center}

{\Large\bf Euclidean correlators at imaginary spatial momentum
\\[2mm] and their relation to the thermal photon emission rate}

\vskip 0.5cm


\end{center}
\vskip 1.5cm

\centerline{\large Harvey B.\ Meyer}
\vskip 1.0cm
\noindent\hspace{-0.25cm} \emph{PRISMA Cluster of Excellence, Institut f\"ur Kernphysik and 
Helmholtz Institut Mainz, Johannes Gutenberg-Universit\"at Mainz, D-55099 Mainz}

\vskip 1.5cm

\centerline{\bf Abstract}\medskip

\noindent 
The photon emission rate of a thermally equilibrated system is
determined by the imaginary part of the in-medium retarded correlator of the
 electromagnetic current transverse to the spatial momentum of the photon. 
In a Lorentz-covariant theory, this correlator can be parametrized by a scalar function
${\cal G}_R(u\cdot {\cal K},{\cal K}^2)$, where $u$ is the fluid
four-velocity and ${\cal K}$ corresponds to the momentum of the
photon. We propose to compute the analytic continuation of ${\cal G}_R(u\cdot
{\cal K},{\cal K}^2)$ at fixed, vanishing virtuality ${\cal K}^2$, to
imaginary values of the first argument, $u\cdot {\cal K}=
i\omega_n$. At these kinematics, the retarded correlator is equal to
the Euclidean correlator $G_E(\omega_n, k=i\omega_n)$, whose first
argument is the Matsubara frequency and the second is the spatial
momentum. The Euclidean correlator, which is directly accessible in
lattice QCD simulations, must be given an imaginary spatial momentum in order to
realize the photon on-shell condition.  Via a once-subtracted
dispersion relation that we derive in a standard way at fixed ${\cal
  K}^2=0$, the Euclidean correlator with imaginary spatial momentum is
related to the photon emission rate.  The relation allows for a more
direct probing of the real-photon emission rate of the quark-gluon plasma 
in lattice QCD than the dispersion relations which have
been used so far, the latter being at fixed spatial photon momentum $k$
and thus involving all possible virtualities of the photon.

\vspace{1.5cm}


\thispagestyle{empty}

\newpage

\clearpage
\pagenumbering{arabic} 
 \section{Introduction}

The electromagnetic radiation emitted by a medium is one of its
important characteristics.  
Under fairly general conditions, the spectrum of emitted photons can be considered a weakly coupled probe of the medium.
Here we will be concerned with a relativistic and thermally equilibrated medium. We will
have the quark-gluon plasma in mind, the high-temperature phase of strongly-interaction matter.
The quark-gluon plasma at a temperature of 200 to 500\,MeV is studied experimentally 
in heavy-ion collisions, and the spectrum of photons produced in the collisions has been measured
at several center-of-mass energies~\cite{Braun-Munzinger:2015hba,Campbell:2017kbo}. 
A closely related observable is the spectrum of detected lepton pairs
($e^+e^-$ or $\mu^+\mu^-$), which are produced via an off-shell (time-like) photon.
As a further important motivation,
the quark-gluon plasma was present in the first microseconds of the Universe.
In addition to photons, various weakly interacting particles may have been produced at that epoch
by similar mechanisms (see \cite{Asaka:2006rw} for the thermal field theory aspects). 
Some of them, such as keV-scale sterile neutrinos (see \cite{Adhikari:2016bei} for a review), 
could constitute (part of) the dark matter in the universe.

The calculation of the photon emission rate by a strongly coupled
medium such as the quark-gluon plasma is a challenging task. The
asymptotic freedom property of QCD implies that weak coupling methods~\cite{Arnold:2001ms,Ghiglieri:2013gia}
become reliable at sufficiently high temperatures. However, the
convergence of the perturbative series at experimentally accessible
temperatures is doubtful. In contrast, the real-time AdS/CFT
correspondence~\cite{Son:2002sd} allows for calculations of transport coefficients and
the photon emission rate at very strong coupling.  These calculations
however cannot to date be performed in QCD, rather they are performed
in theories whose thermal properties are in many instances found to be
similary to those of the quark-gluon plasma~\cite{CasalderreySolana:2011us}. They thus provide a
qualitatively different picture of the high-temperature phase of
non-Abelian gauge theories, mainly characterized by the absence of
quasiparticles. Finally, lattice QCD simulations can deliver
correlation functions related in a known way to the photon emission
rate. Actually determining the latter from the correlation functions
however involves a numerically ill-posed inverse problem related to
the use of the imaginary time (Matsubara) formalism in lattice QCD;
see \cite{Ghiglieri:2016tvj,Brandt:2015aqk,Aarts:2014nba} for recent
calculations of this type.  The correlation functions used so far are
at fixed spatial momentum, and their relation to the production or
absorption of photons involves all possible photon virtualities. This
feature is somewhat unfortunate, because the processes in the medium
leading to the production of a dilepton pair with a large invariant
mass are quite different from those producing a real photon. The
former process occurs already for non-interacting quarks, while, to
leading order in the fine-structure constant, real-photon emission
starts only at O($\alpha_s$), where $\alpha_s=\frac{g^2}{4\pi}$ is the
strong coupling constant.

From a field theory point of view, the photon emission rate of a
thermally equilibrated system is determined by the imaginary part (i.e.\ the spectral function) of
the retarded correlator of the electromagnetic current in the
medium. The retarded correlator can be parametrized by the (spatially)
longitudinal and the transverse Lorentz scalar functions
${\cal G}_R^{T,L}(u\cdot {\cal K},{\cal K}^2)$, where $u$ is the fluid
four-velocity and ${\cal K}$ corresponds to the momentum of the
photon. The first argument thus corresponds to the photon energy in the rest-frame of the fluid.
In this article, we propose to compute the analytic continuation of ${\cal G}_R(u\cdot
{\cal K},{\cal K}^2)$ at fixed, vanishing virtuality ${\cal K}^2$, to
imaginary values of the first argument, $u\cdot {\cal K}=
i\omega_n$. At these kinematics, the retarded correlator is then equal
to the Euclidean correlator $G_E(\omega_n, k=i\omega_n)$, whose first
argument is the Matsubara frequency and the second is the spatial
momentum. The latter must be given an imaginary value in order to
realize the condition ${\cal K}^2=0$.  Via a dispersion relation 
at fixed ${\cal K}^2=0$, the Euclidean correlator with
imaginary spatial momentum is related to the photon emission rate.
One effect of the imaginary spatial
momentum in the Euclidean correlator is to enhance the contribution of
the low-lying spectrum of screening states in each non-static
Matsubara sector, highlighting the importance of understanding precisely that
spectrum.  In QCD, we will show that the low-lying screening states contribute at
O($\alpha_s$) to the imaginary-momentum correlator~\cite{Brandt:2014uda}.

Anticipating our main result, the spatially transverse Euclidean correlator with 
Matsubara frequency $\omega_n$ and imaginary spatial momentum $k=i\omega_n$, denoted $H_E(\omega_n)\equiv G_E(\omega_n,k=i\omega_n)$,
is related by a once-subtracted dispersion relation to the spectral function at vanishing virtuality via
the equations
\be\la{eq:masterE1}
H_E(\omega_n) - H_E(\omega_r) 
 = \int_0^\infty \frac{d\omega}{\pi}\,{\omega}\,\sigma(\omega)\Big[ \frac{1}{\omega^2+\omega_n^2} -  \frac{1}{\omega^2+\omega_r^2}\Big].
\ee
The differential photon emission rate per unit volume of quark-gluon plasma is determined by the spectral function via~\cite{McLerran:1984ay}
\be\la{eq:phora}
{d\Gamma_\gamma(\vec k)} = e^2 \;\frac{d^3k}{(2\pi)^3\,2k}\; \frac{\sigma(k)}{e^{\beta k}-1},
\ee
with $\beta\equiv 1/T$ the inverse temperature.

The ideas involved in arriving at Eq.\ (\ref{eq:masterE1}) are broadly
related to the work \cite{CaronHuot:2008ni}, which considers the 
bremsstrahlung energy loss of high-energy partons moving in the quark-gluon plasma at weak coupling,
and more particularly O($g$) corrections which are shown to be accessible directly from a
computation within the Matsubara formalism. 
These ideas were put into practice in actual lattice calculations\cite{Panero:2013pla}.
Euclidean correlators with an imaginary frequency (i.e.\ momentum in the time
direction) were proposed in \cite{Ji:2001wha} to access for instance
the forward Compton amplitude in a kinematic regime where it is purely
real; see \cite{Chambers:2017dov} for a recent lattice calculation
thereof. The $\eta_c$ and $\pi^0\to\gamma^{(*)}\gamma^{(*)}$
\cite{Dudek:2006ut,Feng:2012ck,Gerardin:2016cqj} transition form
factors were successfully calculated in lattice simulations using
these ideas. In a numerical treatment within the Matsubara formalism, one is however restricted to
real, discrete values of the Euclidean frequency. To achieve a
vanishing photon virtuality, the only option is therefore to use an
imaginary momentum in a spatial direction.

We introduce the relevant notation and relations and  derive Eq.\ (\ref{eq:masterE1}) in section~\ref{sec:deriv}.
We then discuss further aspects of the Euclidean correlator at imaginary spatial momentum and perform tests 
of the dispersion relation in section~\ref{sec:illu}. Final remarks are collected in section~\ref{sec:concl}.

\section{Derivation of Eq.\ (\ref{eq:masterE1})  \label{sec:deriv}}

We begin with some definitions and consider the full set of retarded current-current correlators,
\be\la{eq:G_R}
G_R^{\mu\nu}(u,{\cal K}) = i\int d^4x\; e^{i{\cal K}\cdot x} \theta(x^0)\, \Big\< [{\rm j}^\mu(x),\, {\rm j}^\nu(0)]\Big\>,
\ee
where $\<{\cal O}\> = \frac{1}{\cal Z} {\rm Tr}\{ e^{-\beta u\cdot {\rm P}} {\cal O}\}$ 
denotes the canonical thermal average for a fluid with four-velocity $u$, and the square brackets denote the commutator of two field operators.
The set of operators   ${\rm P}^\mu$, with ${\rm P}^0=H$ the Hamiltonian, is the energy-momentum vector.
Our convention is that ${\rm j}^\mu = \sum_f Q_f\, \bar\psi_f \gamma^\mu \psi_f$, where the Minkowski-space
Dirac matrices satisfy $\{\gamma^\mu,\gamma^\nu\} = 2\eta^{\mu\nu}$ with $\eta^{\mu\nu}={\rm diag}(1,-1,-1,-1)$.
Furthermore, the four-velocity satisfies $u^2=1$ and takes the form $u^\mu = \eta^{\mu 0}$ in the rest frame of the fluid.
Using current conservation, it is convenient to consider the decomposition 
\ba
{\cal G}_R^{\mu\nu}(u,{\cal K}) &=& {\cal P}^{\mu\nu}_L(u,{\cal K})\, {\cal G}_R^L(u\cdot {\cal K},\,{\cal K}^2) 
+ {\cal P}^{\mu\nu}_T(u,{\cal K})\, {\cal G}_R^T(u\cdot{\cal K},\,{\cal K}^2), 
\phantom{\frac{1}{1}}
\\
{\cal P}^{\mu\nu}_T(u,{\cal K}) &=& \eta^{\mu\nu}  -u^\mu u^\nu  
- \frac{({\cal K}^\mu- (u\cdot {\cal K})u^\mu) ({\cal K}^\nu- (u\cdot {\cal K})u^\nu)}{ {\cal K}^2 - (u\cdot {\cal K})^2 },
\\
{\cal P}^{\mu\nu}_L(u,{\cal K}) &=&  \frac{({\cal K}^2 u^\mu - (u\cdot {\cal K})\,{\cal K}^\mu)
({\cal K}^2 u^\nu - (u\cdot {\cal K})\,{\cal K}^\nu)}{{\cal K}^2({\cal K}^2 - (u\cdot {\cal K})^2)}.
\ea
The projector ${\cal P}^{\mu\nu}_T$ projects onto the linear subspace orthogonal to ${\cal K}$ and $u$, and
 \be
{\cal P}^{\mu\nu}_T(u,{\cal K})+{\cal P}^{\mu\nu}_L(u,{\cal K})= \eta^{\mu\nu} - \frac{{\cal K}^\mu {\cal K}^\nu}{{\cal K}^2} 
\equiv {\cal P}^{\mu\nu}({\cal K}) 
\ee
is the projector onto the direction of ${\cal K}^\mu$. In the vacuum, ${\cal G}_R^T={\cal G}_R^L$
and the polarization tensor is proportional to ${\cal P}^{\mu\nu}({\cal K})$.

We consider the linear combination~\cite{Brandt:2017vgl}
\ba
{\cal G}_R(u\cdot {\cal K},\, {\cal K}^2) &=& \Big(2\,\eta_{\mu\nu} - 3\, {\cal P}_{T,\mu\nu}(u,{\cal K})\Big) \;{\cal G}_R^{\mu\nu}(u,{\cal K})
\phantom{\frac{1}{1}}
\\ &=& \Big(2\, {\cal P}_{L,\mu\nu}(u,{\cal K}) - {\cal P}_{T,\mu\nu}(u,{\cal K})\Big) \;{\cal G}_R^{\mu\nu}(u,{\cal K}). 
\phantom{\frac{1}{1}}
\nonumber
\ea
This linear combination vanishes in the vacuum as a combined consequence of Lorentz symmetry and current conservation.
As a consequence, it is ultraviolet finite. 

Preserving the Lorentz covariance in the decomposition of the
polarization tensor exhibits an analogy with the forward,
spin-averaged Compton amplitude of the nucleon\footnote{The analogy is
  useful even though the Compton amplitude is expressed as a
  time-ordered product of operators rather than a retarded
  commutator.}. The latter is also parametrized by two invariant
functions $T_{1,2}(\nu=u\cdot q,q^2)$ of $\nu$, which corresponds to
the photon energy in the frame in which the nucleon is initially at
rest ($u$ is the four-velocity of the nucleon), and of the photon
virtuality $q^2$. In that context, a dispersion relation in the
variable $\nu$ is written for the Compton amplitudes at fixed photon
virtuality; the relevant spectral functions are the structure
functions $F_{1,2}(\nu,q^2)$.  By contrast, in the thermal QCD
context, dispersion relations at fixed spatial photon momentum have
been used, and we review those in the next subsection. We derive a
fixed-virtuality dispersion relation for the case of real photons in
subsection (\ref{sec:q2fixed}).

\subsection{Rest frame of the fluid and dispersion relation at fixed spatial photon momentum \la{eq:rest}}

In the rest frame of the fluid, we write the polarization tensor as $G_R^{\mu\nu}(\omega,\vec k)$.
The ultraviolet-finite linear combination above then reads
\be\la{eq:lincomb2}
{\cal G}_R(u\cdot {\cal K},\, {\cal K}^2) = 
 2G_{R}^{00}(\omega,\vec k) + \left(\delta^{ij} - 3\frac{k^i k^j}{k^2}\right) G_{R}^{ij}(\omega,\vec k)
\equiv G_R(\omega,k^2) .
\ee
with $\omega = u\cdot {\cal K}$, $k^i = {\cal K}^i$ and $k^2\equiv  \vec k\cdot \vec k$.
Current conservation implies
\be
\omega^2 G_R^{00}(\omega,\vec k) = k^i k^j G_R^{ij}(\omega,\vec k),
\ee
so that for light-like kinematics, $\omega=k$, the spatially longitudinal component of the polarization tensor vanishes
and 
\be
G_R(k,k^2) = \left(\delta^{ij} - \frac{k^i k^j}{k^2}\right) G_{R}^{ij}(k,\vec k)
\ee
coincides with the spatially transverse component of the polarization tensor.

We define the spectral function corresponding to the correlator (\ref{eq:lincomb2}) as 
\ba
\rho(\omega,k^2) &=& \int d^4x \;e^{i(\omega x^0 - \vec k\cdot \vec x)} \Big\<2[V^0(x),V^0(0)]
+\left(\delta^{ij} - 3\frac{k^i k^j}{k^2}\right) [V^i(x),V^j(0)]\Big\> 
\nonumber
\\ &=& 2\,{\rm Im}\,G_R(\omega,k^2).
\ea
The dispersive representation of the real part of the retarded correlator at fixed $k^2$ reads 
\be\la{eq:ReGR}
{\rm Re}\,G_{R}(\omega_r,k^2) = \frac{{\rm P}}{\pi} \int_0^\infty d\omega \,\omega\; \frac{\rho(\omega,k^2)}{\omega^2-\omega_r^2},
\ee
where ${\rm P}$ indicates that the principal value integral should be taken.
We have made use of the fact that $\rho(\omega,k)$ is an odd function of the frequency $\omega$.
Due to the fact that $\rho(\omega,k^2)\sim k^2/\omega^4$ at large $\omega$ for fixed $k^2$~\cite{Brandt:2017vgl},
no subtraction terms are required to make the dispersive integral convergent.
We also define the spectral function at fixed, vanishing photon virtuality by
\be
\sigma(\omega) = \rho(\omega,k^2=\omega^2),
\ee
in terms of which the photon emission rate is obtained according to Eq.\ (\ref{eq:phora}).
The  dispersive representation (\ref{eq:ReGR}) can be derived in two ways: the first is to use the 
spectral representation of the retarded correlator, obtained by inserting a complete set of states inside
the operator product $V^\mu(x) V^\nu(0)$ (see for instance \cite{Meyer:2011gj}); the other is to inspect the analytic properties
of the retarded and the advanced correlators (the latter is obtained by replacing $\theta(x^0)$ by $-\theta(-x^0)$ in Eq.\ (\ref{eq:G_R})).
In this derivation, Eq.\ (\ref{eq:ReGR}) is the result of expressing an analytic function via Cauchy's theorem.
This type of dispersion relation, namely at fixed spatial momentum, hold both for relativistic and non-relativistic theories.
However, causality has stronger consequences in a relativistic theory:
the commutator of two local field operators vanishes at all spacelike separations.
This allows one to obtain a dispersion relation in the photon energy at fixed photon virtuality ${\cal K}^2 \equiv \omega^2-k^2$.

\subsection{Dispersion relation at fixed photon virtuality \la{sec:q2fixed}}

\begin{figure}
\centerline{\includegraphics[width=0.4\textwidth]{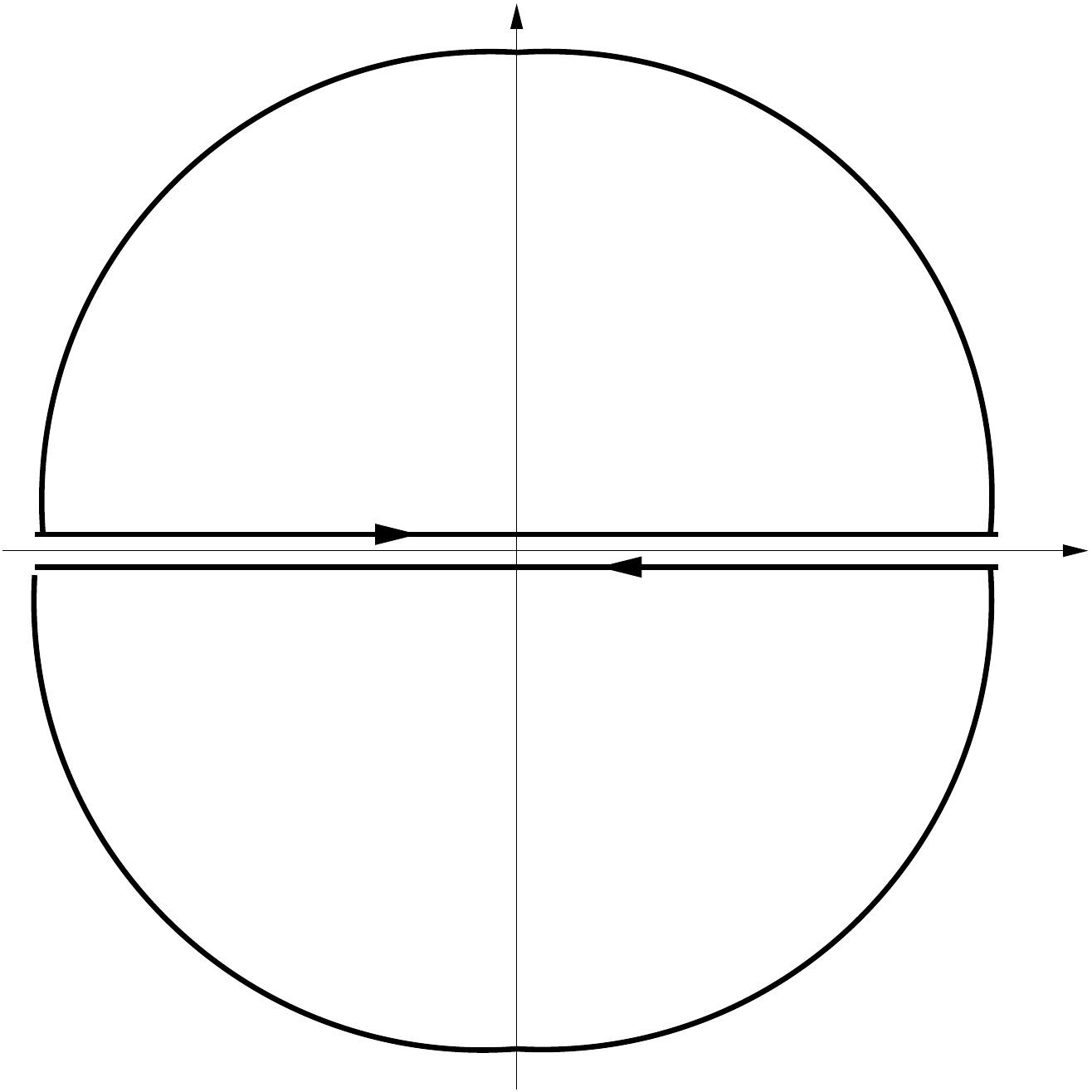}}
\caption{\label{fig:contour}
The contour in the complex $\omega$ plane used for the dispersive representation of the function $H(\omega)$ via Cauchy's theorem.
In the upper half plane, $H(\omega)$ coincides with the retarded correlator $H_R(\omega)$. The variable $\omega$ represents the photon energy
in the rest frame of the fluid.}
\end{figure}

We recall the standard derivation of the dispersion relation at fixed, vanishing photon virtuality~\cite{Weinberg:1995mt}.
In view of the vanishing of the commutator outside the light-cone, 
\be\la{eq:relcausal}
[j^\mu(x),j^\nu(0)] = 0 \quad {\rm for}\quad x^2<0,
\ee
the retarded correlator $H_R(\omega) \equiv G_R(\omega,\omega)$ at lightlike momentum is analytic for ${\rm Im}\,(\omega)>0$.
Likewise, the advanced correlator $H_A(\omega) \equiv G_A(\omega,\omega)$ is analytic for ${\rm Im}\,(\omega)<0$.
The idea is then to define the function 
\be
H(\omega) = \left\{ \begin{array}{ll} H_R(\omega) & {\rm Im}\,(\omega)>0 \\ H_A(\omega) & {\rm Im}\,(\omega)<0
\end{array}\right.,
\ee
which is analytic everywhere, except for a discontinuity on the real axis.
The discontinuity is given by 
\be
H(\omega+i\epsilon) - H(\omega-i\epsilon) = H_R(\omega) - H_A(\omega) = i \sigma(\omega),
\ee
which corresponds to the Fourier transform of the commutator over the entire time axis.
We then represent the correlator as a contour integral in the complex plane 
via Cauchy's theorem. The chosen contour  is displayed in Fig. (\ref{fig:contour}).
Thus the function $H(\omega)$ just above the real axis, where it coincides with $H_R(\omega)$,
can be obtained as a contour integral,
\be
H_R(\omega) = H_R(\omega_r)  + \int_{-\infty}^\infty \frac{d\omega'}{2\pi}
\; \sigma(\omega')\Big[ \frac{1}{\omega'-\omega-i\epsilon} - \frac{1}{\omega'-\omega_r-i\epsilon}\Big].
\ee
We have anticipated that a subtraction is necessary to make the representation convergent.
Indeed, in the examples considered in section \ref{sec:illu}, $\sigma(\omega)$ grows like $\omega^{\gamma}$
with $\frac{1}{2}\leq \gamma < 1$ at large frequencies.
Taking the real part, we obtain the  representation 
\be
{\rm Re}\,H_R(\omega) = {\rm Re}\,H_R(\omega_r)  + {\rm P}\int_{-\infty}^\infty \frac{d\omega'}{2\pi}
\; \sigma(\omega')\Big[ \frac{1}{\omega'-\omega} - \frac{1}{\omega'-\omega_r}\Big]
\ee
of the real part of $H_R(\omega)$ in terms of a principal value integral over the imaginary part of $H_R(\omega)$.
Using the fact that $\sigma(\omega)$ is odd, the dispersion relation can be rewritten
\be\la{eq:masterR}
{\rm Re}\,H_R(\omega) = {\rm Re}\,H_R(\omega_r)  + {\rm P}
\int_{0}^\infty \frac{d\omega'\;\omega'}{\pi}
\; \sigma(\omega')\Big[ \frac{1}{\omega'{}^2-\omega^2} - \frac{1}{\omega'{}^2-\omega_r^2}\Big].
\ee
This is the master relation that allows one to probe the photon emission rate via 
the real part of the retarded correlator.

At this point, one may wonder whether the subtraction point $\omega_r$ can be sent to zero.
For this to be useful, the behavior of $H_R(\omega)$ must be known in this regime.
According to hydrodynamics, the longitudinal part of the retarded correlator 
is dominated in the limit of small $\omega,k$ by the diffusion pole (see e.g.\ \cite{Meyer:2011gj}), 
\ba
G_R^{00}(\omega,k) &=& \frac{(Dk^2)^2 + i\omega \,Dk^2}{\omega^2+(Dk^2)^2} \,\chi_s,
\\
\frac{k^i k^j}{k^2} G_R^{ij}(\omega,k) &=& \frac{\omega^2}{k^2} \,G_R^{00}(\omega,k),
\ea
where $\chi_s$ is the static susceptibility and $D$ the diffusion coefficient.
Since the combination $\Big(\delta^{ij} - 3\frac{k^i k^j}{k^2}\Big) G_R^{ij}(\omega,k)$ vanishes in the limit $k\to 0$,
the spatially transverse part must be given in that limit by
\be
\lim_{k\to0} \Big(\delta^{ij} - \frac{k^i k^j}{k^2}\Big) G_R^{ij}(\omega,k) 
=  2  i\omega \,D \,\chi_s \qquad (\omega{\rm~fixed,~}\omega\lesssim D^{-1}).
\ee
This equation is the content of the Kubo formula for the diffusion coefficient. We note that the real part
of the retarded correlator vanishes in the limit $k\to 0$ at fixed $\omega$.
What happens generically in the light-like limit $(\omega=k)\to 0$ is not entirely clear to us.
As described in section \ref{sec:ads}, in this limit $H_R(\omega)$ tends to zero 
in the ${\cal N}=4$ SYM theory at strong coupling and large-$N_c$,
while $H_R(\omega)$ is simply constant and non-vanishing for $\omega\neq 0$ in the theory of non-interacting quarks.
A dedicated study of this limit in kinetic theory along the lines of~\cite{Hong:2010at} could be illuminating.

\subsection{Euclidean correlator at imaginary spatial momentum}

In the Matsubara formalism, one considers the momentum-space correlators
\be
G_{E,\mu\nu}(\omega_n,\vec k) = \int_0^\beta dx_0\; e^{i\omega_n x_0} \int d^3x \; e^{i\vec k\cdot \vec x}\; \< J_\mu(x) J_\nu(0)\>,
\ee
where $\omega_n=2\pi Tn$ and $J_\mu(x) = \sum_f Q_f  \bar\psi_f \gamma_\mu^{_{\rm E}} \psi_f$, the Euclidean Dirac matrices satisfying 
$\{\gamma_\mu^{_{\rm E}},\gamma_\nu^{_{\rm E}}\}=2\delta_{\mu\nu}$, i.e.\ the spatial gamma matrices are Hermitian
as opposed to anti-Hermitian. 
The Euclidean correspondent of the linear combination (\ref{eq:lincomb2}) then takes the form
\be
G_E(\omega_n,k^2) = 2G_{E,00}(\omega_n,\vec k) -\left(\delta_{ij} - 3\frac{k_i k_j}{k^2}\right) G_{E,ij}(\omega_n,\vec k),
\ee
so that~\cite{Meyer:2011gj}
\be
G_E(\omega_n,k^2) = G_R(i\omega_n,k^2), \qquad n> 0.
\ee
Extending the function to imaginary spatial momentum, we have 
\be
H_E(\omega_n)\equiv G_E(\omega_n,-\omega_n^2) = G_R(i\omega_n,-\omega_n^2) = H_R(i\omega_n).
\ee
This continuation is possible straightforwardly if the integral over the spatial volume still converges.
Thus translating Eq.\ (\ref{eq:masterR}) above into Euclidean notation, we have 
\be\la{eq:masterE2}
H_E(\omega_n) -H_E(\omega_r) 
 = \int_0^\infty \frac{d\omega}{\pi}\;{\omega}\,\sigma(\omega)\Big[ \frac{1}{\omega^2+\omega_n^2} -  \frac{1}{\omega^2+\omega_r^2}\Big],
\qquad n,r\neq 0.
\ee
We expect this relation to be useful because the left-hand side can be computed by standard techniques in lattice QCD.
The reference frequency $\omega_r$ would be set to $\omega_1=2\pi T$.
The left-hand side thus probes the real-photon spectral function, which for typical thermal photon energies is of order $\alpha_s$
at weak coupling.
Indeed, we shall review in section \ref{sec:freequarks} that the left-hand side vanishes for non-interacting quarks.
Mathematically speaking, by Carlson's theorem of complex analysis 
the knowledge of $H_E(\omega_n)$ for all $n\geq n_0$, where $n_0$ is any natural number, uniquely determines
the spectral function $\sigma(\omega)$~\cite{Cuniberti:2001hm,Meyer:2011gj,Ferrari:2016snh}.
Numerically, this complete knowledge cannot be achieved. Nonetheless it is very interesting to compute a quantity
in lattice QCD which is directly sensitive to interactions in the quark-gluon plasma and is directly related to the 
observable photon emission rate.

\section{Further aspects and tests of the dispersive representation\la{sec:illu}}

In this section, we illuminate further aspects of the Euclidean
correlator at light-like momenta.  We begin by discussing the infrared
contributions to the correlator in terms of non-static screening
states.  In the second subsection, we comment on a technical issue
that arises due to the fact that the lattice regularization breaks the
Lorentz symmetry.  We then perform tests of the dispersion relation in
the theory of non-interacting quarks and in the case of correlators
obtained via AdS/CFT methods. The latter correspond to a strongly coupled
plasma.

\subsection{Representation of $H_E(\omega_n)$ through non-static screening states}

At light-like kinematics, only the spatially transverse part of the polarization tensor contributes to
$H_E(\omega_r)$. Thus, we consider the transverse `screening' correlator,
\be
G_E^T(\omega_r,x_3) \equiv -\sum_{i=1}^2 \int_0^\beta \!dx_0\; e^{i\omega_r x_0}\!\int dx_1dx_2\; \<J_i(x)J_i(0)\> 
\stackrel{x_3\neq 0}{=} \sum_{n=0}^\infty |A_n^{(r)}|^2\, e^{-E_n^{(r)}|x_3|}
\ee
which has a representation in terms of the energies $E_n^{(r)}$ and amplitudes $A_n^{(r)}$ of non-static screening states.
The lowest level $E_0^{(r)}$ in the Matsubara sector $\omega_r$ determines the asymptotic spatial fall-off of the correlator.
This low-lying screening spectrum has been studied in~\cite{Brandt:2014uda} both at weak coupling and in lattice QCD.
In terms of these levels, the correlator at light-like kinematics $H_E(\omega_r)$ can be written as 
\be
H_E(\omega_r)= \int_{-\infty}^\infty dx_3\; G_E^T(\omega_r,x_3)\; e^{\omega_r x_3}
= 2\sum_{n=0}^\infty |A_n^{(r)}|^2\; \frac{E_n^{(r)}}{E_n^{(r)\,2} - \omega_r^2} + {\rm \;c.t.},
\ee
where ${\rm \;c.t.}$ stands for contact terms originating from the region around $x_3=0$.
Given that both $A_0^{(r)}$ and $E_0^{(r)}-|\omega_r|$ are of order $g^2$ (with $g$ the QCD gauge coupling),
the low-lying spectrum makes an order $g^2$ contribution to $H_E(\omega_r)$.
This helps explain the connection observed in~\cite{Brandt:2014uda} between non-static screening masses and the
Landau-Pomeranchuk-Migdal (LPM) resummed contributions to the photon emission rate as computed in~\cite{Aurenche:2002wq}.
In particular, the lattice results~\cite{Brandt:2014uda} for $E_0^{(1)}$ were found to be in line with the prediction based on an effective
theory (EQCD), while the amplitude $|A_0^{(1)}|^2$ was found to be a factor $(7.7\pm2.9)$ larger from the lattice simulation
than its EQCD prediction. This would indicate a much stronger infrared contribution to the photon emission rate.

It is worth mentioning that $E_0^{(r)}-|\omega_r|$ cannot be negative. An argument why it must be so goes as follows.
The position-space Euclidean correlator of any two local currents $A$ and $B$ can be represented by a Fourier series,
\[
\hat G_E(x)= T\sum_{\ell\in\mathbb{Z}} e^{-i\omega_\ell x_0} \; \tilde G_E(\omega_\ell,\vec x).
\]
Then the Wightman correlator for spacelike separations, $t^2-\vec x^2<0$, being the analytic continuation of $\hat G_E(x)$ 
back to real time, is given by the Fourier coefficients according to
\[
G_>(t,\vec x) \equiv \frac{1}{Z}{\rm Tr}\{e^{-\beta H} A(t,\vec x) B(0)\} 
= T \sum_{\ell\in\mathbb{Z}} e^{\omega_\ell t} \;\tilde G_E(\omega_\ell,\vec x).
\]
Thus if $G_E(\omega_\ell,\vec x)$ fell off as $e^{-E_0^{(\ell)}|\vec x|}$ with $E_0^{(\ell)}<|\omega_\ell|$ for one particular
integer $\ell$, then in the space-like direction  $t= |\vec x|(1-\epsilon E_0^{(\ell)}/|\omega_\ell|)$ with $\epsilon$ chosen such that 
$|\omega_\ell|> (1+\epsilon)E_0^{(\ell)}$, the Wightman correlator would diverge exponentially at long distances\footnote{We are assuming here 
that the other Matsubara sectors cannot conspire to cancel this growing exponential.}, 
which is clearly unphysical. The property $E_0^{(r)}-|\omega_r|>0$ ensures that $H_E(\omega_\ell)$ is well-defined in the infrared. 
Both weak-coupling and lattice calculations indeed find that $E_0^{(r)}-|\omega_r|$ 
is strictly positive~\cite{Brandt:2014uda}. Non-interacting massless quarks constitute a limiting case, where 
the low-lying states form a $q\bar q$ continuum in a $p$-wave and a direct calculation~\cite{Brandt:2014uda} shows for instance that 
$G_E^T(\omega_1,x_3) \sim e^{-\omega_1 |x_3|}/(x_3)^2$ at long distances. The additional suppression by the second inverse power of $x_3$
suffices to make $H_E(\omega_1)$ infrared-safe.

\subsection{Lorentz symmetry and the correlator $H_E(\omega_n)$ on the lattice \la{sec:LosyLat}}

An ultraviolet issue in the calculation of the light-like correlator
$H_E(\omega_n)$ arises on the lattice.  In the continuum,
$G_E(\omega_n,k^2)$, and hence
$H_E(\omega_n)=G_E(\omega_n,-\omega_n^2)$ vanish identically in the
vacuum.  This property however relies crucially on Lorent
symmetry. Since the lattice regulator breaks this symmetry,
short-distance contributions on the lattice can spoil the continuum
limit of $H_E(\omega_n)$.

One safe remedy against this problem is to explicitly subtract from
the thermal lattice correlator $H_E(\omega_n)$ the corresponding
vacuum lattice correlator $H_{E,{\rm vac}}(\omega_n)$, obtained at the
same bare parameters (quark masses and gauge coupling). This is
appropriate, since the vacuum correlator vanishes in a
Lorentz-symmetry-preserving regularization. The subtraction has the
effect that, in an operator-product expansion analysis of the
short-distance contributions to $H_E(\omega_n)$ performed in lattice
perturbation theory, the contribution of the unit operator cancels out
and the remaining short-distance singularities of the vector
correlator are integrable. In this way, one is not relying on Lorentz
symmetry for the cancellation of short-distance divergences, but only 
on the absence of dimension-two gauge-invariant operators in the theory,
which is guaranteed by the exact SU(3) gauge symmetry of lattice QCD.
With the vacuum-subtraction in place, and
with an on-shell O($a$) improved lattice discretization of the action
and the vector currents~\cite{Luscher:1996ug}, we expect the left-hand
side of the master relation (\ref{eq:masterE2}) to approach its
continuum limit with O($a^2$) corrections.

\subsection{Test of the dispersion relation for non-interacting quarks\label{sec:freequarks}}

In this subsection, we use a Euclidean notation, set the quark electric charge to unity
and consider the case of $N_c$ non-interacting Dirac fermions of mass $m$.  The polarization tensor is then given  by
\be\la{eq:Pi1}
\Pi_{\mu\nu}(K) = - N_c T \sum_{P_0\in\Gamma_F}\int \frac{d^3P}{(2\pi)^3} \,{\rm Tr}
\left\{\gamma^{_{\rm E}}_\mu\frac{-i/\!\!\!\!P+m}{P^2+m^2}\gamma_\nu^{_{\rm E}}\frac{-i/\!\!\!\!P-i/\!\!\!\!K+m}{(P+K)^2+m^2} \right\}
\ee
where $\Gamma_F$ is the set of fermionic Matsubara frequencies, $P_0= (2n+1)\pi T$, $n\in\mathbb{Z}$.
We treat separately the `vacuum' and the `matter' contributions, 
$\Pi_{\mu\nu}(K) = \Pi_{\mu\nu}^{\rm vac}(K) + \Pi_{\mu\nu}^{\rm mat}(K)$.
The vacuum contribution has the generic form
\be\la{eq:VacPolTens}
\Pi^{\rm vac}_{\mu\nu}(K) = (K_\mu K_\nu - \delta_{\mu\nu}K^2) \Pi(K^2),
\ee
the virtuality dependence of $\Pi(K^2)$ being given at one loop by
\be
\Pi(K^2)-\Pi(0) = \frac{N_c}{2\pi^2}\int_0^1 dx \, x(1-x)\log\left[1+x(1-x)K^2/m^2\right].
\ee
The vacuum polarization $\Pi(K^2)$ contains a logarithmic divergence which cancels in the subtraction above.
Note however that the form (\ref{eq:VacPolTens}), which follows from Lorentz invariance and current conservation, 
implies that the spatially transverse part, e.g. $\Pi^{\rm vac}_{11}(K)$ with $K=(k_0,0,0,k)$, vanishes
for $k_0=\pm i  k$.

For the matter part, there are two independent scalar functions to be calculated.
For two sets of components, we can write~\cite{Kapusta:2006pm}
\be\la{eq:PiX}
\Pi_{X}^{\rm mat}(K) = \frac{4N_c}{\pi^2} {\rm Re}\int_0^\infty {p^2dp}\, \frac{n_F(E_{\vec p})}{E_{\vec p}}
\, J_X(k_0,k,p),
\ee
with $X=00$ or $X=\mu\mu$ and 
\ba
J_{00}(k_0,k,p) 
 &=& \frac{1}{2} + \frac{4E_{\vec p}^2 - 4iE_{\vec p}k_0 - k_0^2 - k^2}{8pk}\,
{\rm Log},
\\
J_{\mu\mu}(k_0,k,p) &=& 
1+\frac{2m^2-k_0^2-k^2}{4kp} \;{\rm Log},
\\
{\rm Log} & \equiv & \log\left[\frac{k_0^2+ k^2 + 2pk + 2iE_{\vec p}k_0}{k_0^2+ k^2 - 2pk +2iE_{\vec p}k_0} \right].
\ea
In Eq.\ (\ref{eq:PiX}), the operation ${\rm Re}(f(k_0))$ should be interpreted as the average 
$(f(k_0)+f(-k_0))/2$.
From these components, the linear combination of interest can be expressed as 
\ba
G_E(k_0,k) &=& -\Pi_{\mu\mu}(k_0,\vec k) + 3 \left(1+\frac{k_0^2}{k^2}\right)\Pi_{00}(k_0,\vec k).
\\ &=& -\Pi_{\mu\mu}^{\rm mat}(k_0,\vec k) + 3 \left(1+\frac{k_0^2}{k^2}\right)\Pi_{00}^{\rm mat}(k_0,\vec k).
\ea
In particular, setting $k=ik_0$, we obtain a result independent of $k_0$,
\be
H_E(k_0) = G_E(k_0,ik_0)= \frac{-4N_c}{\pi^2} \int_0^\infty dp\,p^2\, \frac{n_F(E_{\vec p})}{E_{\vec p}} \stackrel{m=0}{=} -\frac{N_c}{3}T^2.
\ee
Thus in the free theory, both sides of Eq.\ (\ref{eq:masterE2}) vanish.
We note that for general mass $m$, the expression above has an interpretation as the vacuum-subtracted thermal chiral condensate,
\be
\frac{-4N_c}{\pi^2} \int_0^\infty dp\,p^2\, \frac{n_F(E_{\vec p})}{E_{\vec p}} =
 - 2\frac{\partial}{\partial m} (\<\bar\psi \psi\>_T - \<\bar\psi\psi\>_{\rm vac}) .
\ee
It would be interesting to see whether the relation between $H_E(k_0)$ and the chiral condensate
generalizes to the interacting theory.
It is important to note that the limits $k_0\to0$ and $k\to0$ do not commute. In particular,
\be
\lim_{k\to0} G_E(0,k) \stackrel{m=0}{=} +\frac{N_c}{3}T^2 
\ee
differs from $ \lim_{k_0\to0} H_E(k_0)$.

\subsection{Large-$N_c$ ${\cal N}=4$ SYM at infinite 't Hooft coupling\la{sec:ads}}

The retarded correlator of two vector currents in the large-$N_c$ ${\cal N}=4$ super-Yang-Mills theory at infinite  
't Hooft coupling $\lambda$ can be computed using the AdS/CFT real-time prescription~\cite{Son:2002sd}.
For the purpose of testing the dispersive representation (\ref{eq:masterE2}) in an interacting theory, 
we have obtained both the real and the imaginary part of the retarded
correlator by solving ordinary differential
equations~\cite{Kovtun:2005ev,CaronHuot:2006te} for the longitudinal and transverse
electric field.  The imaginary part of the retarded correlator for light-like momenta 
has an analytic solution~\cite{CaronHuot:2006te} in terms of the hypergeometric function $_2F_1$,
\be\la{eq:hyperg}
\frac{\sigma(\omega)}{\chi_s} = 
\frac{w}{\left| \, _2F_1\left(1-\left(\frac{1}{2}+\frac{i}{2}\right)
   w,\left(\frac{1}{2}-\frac{i}{2}\right) w+1;1-i w;-1\right)\right| {}^2},
%
\quad w = \frac{\omega}{2\pi T}.
\ee

Here we find numerically that 
\be
\lim_{\omega\to0} \,{\rm Re}\; H_R(\omega) = 0,
\ee
so that the dispersion relation (\ref{eq:masterR}) can be written in the simpler form
\be\la{eq:masterR0}
{\rm Re}\,H_R(\omega) = \frac{\omega^2}{\pi}\; {\rm P} \int_{0}^\infty \frac{d\omega'}{\omega'} \frac{\sigma(\omega')}{\omega'{}^2-\omega^2}\,,
\ee
or in the Euclidean version 
\be\la{eq:masterE0}
H_E(\omega_n) = -\frac{\omega_n^2}{\pi} \int_{0}^\infty \frac{d\omega'}{\omega'} \frac{\sigma(\omega')}{\omega'{}^2+\omega_n^2}\,.
\ee
In the following, for the purpose of performing numerical checks, we divide both sides by the static susceptibility
$\chi_s = \frac{N_c^2 T^2}{8}$, thus rendering them dimensionless.
\begin{itemize}
\item We checked Eq.\ (\ref{eq:masterE0}) by computing the left-hand side
numerically from the differential equations and finding
$(-1.343,\;-2.230)$ for the cases $n=1$ and $n=2$; on the other hand,
computing the right-hand side integral numerically using
Eq.\ (\ref{eq:hyperg}), we obtain the same answer to the quoted number of digits.
\item Similarly, we checked Eq.\ (\ref{eq:masterR0}) in the same way
and find, for $\omega=\pi T$, $-0.2543(1)$ for the right-hand side (using the standard $\epsilon$-prescription 
to perform the principal-value integral), and $-0.2544(1)$ for the left-hand side.
For $\omega=2\pi T$, we find $-0.6030(1)$ for the left- and right-hand side.
\item We have also checked the dispersive representation at fixed spatial momentum, Eq.\ (\ref{eq:ReGR}).
For instance, at $\omega=\pi T$, the right-hand side value of $-0.2544(1)$  is reproduced 
by inserting the  spectral function $\rho(\omega',k=\pi T)$ numerically determined for all $\omega'$
into the integral, for which we obtain $-0.252(3)$. In this case the precision of the test is only at the one-percent level, owing to
the cumulated uncertainty of inserting a numerically determined function into a principal-value integral.
\end{itemize}


\section{Conclusion\label{sec:concl}}

We have derived and tested a dispersion relation at fixed virtuality for Euclidean correlators
at imaginary spatial momentum in terms of the photon
emission rate of thermal QCD matter, Eqs.\ (\ref{eq:masterE1}--\ref{eq:phora}). 
The dispersive variable is the photon energy in the rest frame of the fluid.
We have mainly the photon rate of the quark-gluon plasma in mind, but 
the relation applies equally well to the low-temperature phase of QCD.
Dispersion relations formulated at fixed spatial momentum
(e.g.\ Eq.\ (\ref{eq:ReGR})) only exploit the non-relativistic version
of causality, namely the fact that the retarded correlator
(\ref{eq:G_R}) is analytic for ${\rm Im}(\omega)>0$ for any fixed
spatial momentum $k$; those at fixed virtuality ${\cal K}^2=0$ make
use of the stronger relativistic causality property, see
Eq.\ (\ref{eq:relcausal}). 

It remains to be seen in practice how well the Euclidean correlators
at imaginary spatial momentum can be controlled (a) at long distances,
where perhaps a larger spatial extent in the direction of the momentum
will have to be used; and (b) at short distances, where a subtraction
of the Lorentz-symmetry breaking effects of the lattice must be
performed, as described in subsection \ref{sec:LosyLat}. But if these
correlators can be determined reliably, comparisons with weak-coupling
predictions of $\sigma(\omega)$ promise to be extremely interesting, since
the latter involve sophisticated
resummations~\cite{Arnold:2001ms,Aurenche:2002wq,Ghiglieri:2013gia}.
Also, if one attempts a numerical inversion of the dispersion relation
(\ref{eq:masterE1}) for $\sigma(\omega)/\omega$, an important feature of the latter function is that it is expected to have a very 
mild $\omega$ dependence in the quark-gluon plasma (see e.g.\ Fig.\ 11 in \cite{Brandt:2017vgl}), 
except at weak coupling when one enters the $\omega\lesssim D^{-1}$ regime, where $D$ is the diffusion coefficient; however, 
the Lorentzian kernel appearing in (\ref{eq:masterE1}) suppresses this region.
This feature is helpful in reconstructing $\sigma(\omega)$ for $\omega\approx 2\pi T$,
while it means that the soft-photon emission rate remains difficult to probe using Euclidean correlators,
a point that was already noticed~\cite{Ghiglieri:2016tvj,Brandt:2017vgl} in the fixed-$k$
dispersion relation.

Finally, it is likely that Euclidean correlators at imaginary momentum have
further interesting applications in lattice QCD, including in
zero-temperature physics.

\section*{Acknowledgments}

I thank my collaborators B.B.\ Brandt, M.\ C\`e, A.\ Francis, M.T.\ Hansen, T.\ Harris, D.~Robaina, S.~Sharma, A.\ Steinberg 
and K.\ Zapp in the endeavor of extracting real-time properties of QCD from lattice simulations
for many stimulating discussions over the past few years.
This work is part of a project that has received funding from the
European Research Council (ERC) under the European Union's Horizon
2020 research and innovation programme under grant agreement No.\ 771971--SIMDAMA.
Support by the Deutsche Forschungsgemeinschaft via grant No.\ ME 3622/2-2 is greatfully
acknowledged.
\bibliographystyle{JHEP}
\bibliography{/Users/harvey/BIBLIO/viscobib}
\end{document}